\documentclass[showpacs,twocolumn,amsmath,amssymb,pre]{revtex4}

\usepackage{graphicx,bm,epsfig}

\begin{document}

\title{Heterogeneous micro-structure of percolation in sparse networks}

\author{Reimer K\"uhn$^{1}$ and Tim Rogers$^{2}$ }

\affiliation
{
	$^{1}$ Department of Mathematics, King's College London, Strand, London WC2R 2LS, UK\\
	$^{2}$  Centre for Networks and Collective Behaviour, Department of Mathematical Sciences, University of Bath, Bath, BA2 7AY, UK.\\
}

\begin{abstract}
We examine the heterogeneous responses of individual nodes in sparse networks to the random removal of a fraction of edges. Using the message-passing formulation of percolation, we discover considerable variation across the network in the probability of a particular node to remain part of the giant component, and in the expected size of small clusters containing that node. In the vicinity of the percolation threshold, weakly non-linear analysis reveals that node-to-node heterogeneity is captured by the recently introduced notion of non-backtracking centrality. We supplement these results for fixed finite networks by a population dynamics approach to analyse random graph models in the infinite system size limit, also providing closed-form approximations for the large mean degree limit of Erd\H{o}s-R\'enyi random graphs. Interpreted in terms of the application of percolation to real-world processes, our results shed light on the heterogeneous exposure of different nodes to cascading failures, epidemic spread, and information flow. 
\end{abstract}


\maketitle

\section{Introduction}
Since the very begining of the modern fascination with networked systems, researchers have been interested in questions of propagation. Across many applications, the analysis of bond percolation provides a simple framework with which to analyze the capability of a network to transmit information, disease, influence, or failure \cite{Grassberger1983,Callaway2000,Newman2002,Watts2002}. Early work in this area mainly concentrated on understanding the global properties of percolation in the ensemble average of randomly generated model networks. Surprisingly, detailed results for single instances of fixed networks have only been available relatively recently \cite{Karrer2014, Hamilton2014}, and very little is known exactly about the responses of individual nodes \cite{Rogers2015}.

We consider bond percolation for fixed networks defined as follows: starting from an arbitrary large (connected) network, we evaluate each edge independently, keeping it with probability $\rho$ and deleting it with probability $1-\rho$. The largest connected component remaining after this random edge removal process is referred to as the \textit{percolating cluster} or \textit{giant component}; write $S$ for its size measured as a fraction of the total number $N$ of nodes in the network. For large sparse networks it was shown in \cite{Karrer2010, Karrer2014} that this quantity can be computed to close approximation using a message-passing protocol. 

Knowledge of $S$ gives global information about the robustness of a network to attack or infection; in particular, there is a critical value of $\rho$ below which no percolating cluster survives in the thermodynamic limit, and $S=0$. One of the main results of \cite{Karrer2014, Hamilton2014} was to identify the percolation threshold $\rho_c$ as the reciprocal of the largest eigenvalue of the \textit{non-backtracking} (or \textit{Hashimoto}) matrix that encodes the relationship between variables in the message-passing equations. The eigenstructure of this matrix has received considerable attention recently, having been proposed as an efficient tool for both network clustering \cite{Krzakala2013} and centrality analysis \cite{Martin2014}. 
\begin{figure}
\includegraphics[width=240pt, trim=10 30 10 10]{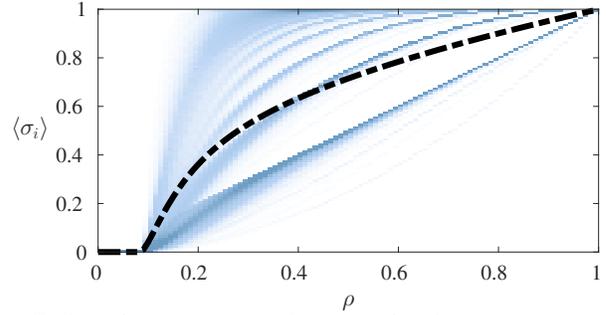}
\caption{Micro-structure of percolation in a sample network with 62,586 nodes taken from the \textit{gnutella} file sharing platform \cite{Snap2014}.  Each vertical slice of the density plot shows the distribution $\varphi(s)$ of probability to appear in the percolating cluster for given edge occupation probability $\rho$. The think dashed line shows the expected size of the percolating cluster $S$, which is equal to the mean of $s$ under $\varphi$.}
\label{gnutella}
\end{figure}

In this article we will be concerned with more detailed questions about the typical outcomes for individual nodes in the network, when averaged over many instances of the percolation process. For a given random instantiation of percolation, write $\sigma_i=1$ if node $i$ appears in the largest connected component, and $\sigma_i=0$ if not. Taking the ensemble average of this variable yields the probability $\langle \sigma_i\rangle$ for node $i$ to appear in the percolating cluster. Heterogeneity in the responses of individual nodes to percolation is captured by the empirical distribution of $\langle \sigma_i\rangle$, defined as 
\begin{equation}
\varphi(s)=\frac{1}{N}\sum_i\delta(s-\langle \sigma_i\rangle)\,.
\label{defpi}
\end{equation}
Notice that the total fractional size of the percolating cluster is given by the mean of $\varphi$, that is, $S=\int s\,\varphi(s)\,\text{d}s$. When node $i$ does not appear in the percolating cluster, write $n_i$ for the size of the component it belongs to, and $\langle n_i\rangle$ for the average over many instances. The node average $\frac{1}{N}\sum_i\langle n_i\rangle$ was again analysed for finte networks in \cite{Karrer2014}, and previously results for the distribution of finite cluster sizes in large random graphs was presented in \cite{Callaway2000}. The empirical distribution small clusters is defined analogously to $\varphi$ in Eq.~\eqref{defpi}; 
\begin{equation}
\psi(n)=\frac{1}{N}\sum_i\delta(n-\langle n_i\rangle)\,.
\end{equation}
It turns out that many networks exhibit extreme differences between nodes in both $\langle \sigma_i\rangle$ and $\langle n_i\rangle$, which are not well-represented by the average value of percolation probability or small cluster size. Figure~\ref{gnutella} showsn an illustraive example for the probability to appear in the percolating cluster of nodes in a real-world dataset. This behaviour was previously observed in the particular context of epidemic spreading on networks, and exploited in \cite{Moreno2002} to formulate and analyse a heterogeneous dynamic mean-field approximation of epidemic spreading. Within that approximation the infection probability of nodes is postulated to depend \textit{only} on the degrees of individual nodes, an assumption which allows of a self-consistent solution. This degree-based approximation has become a mainstay in the analysis of epidemic spreading on networks, including in particular also in the search for optimal vaccination strategies; see \cite{Pastor2015} for a recent review. The heterogeneity was again observed within a cavity formulation of the problem \cite{Rogers2015} which for the analysis of SIR dynamics allows an exact mapping on bond-percolation. In that work it was found that node degrees play a dominant role in the behaviour of $\langle \sigma_i\rangle$ near $\rho=1$, but also that the picture becomes much more complex near $\rho_c$.  

In this paper we explore these issues in detail. After recapping the message passing formulation in the next section we move on in section \ref{weak} to consider node variability in fixed finite networks in the neighbourhood of the percolation transition. In this regime we apply a weakly non-linear analysis to compute the probability of a node to appear in the percolating cluster, and the expected size of non-percolating clusters containing that node. In fact, we will show that the measure of non-backtracking centrality proposed in \cite{Martin2014} determines the leading-order behaviour of both $\langle \sigma_i\rangle$ and $\langle n_i\rangle$ near percolation. This gives and interesting physical interpretation to a quantity that was originally devised for purely practical reasons. In section \ref{popdyn} we move on to use a population dynamics approach compute the distributions of node percolation probability and expected small component size in the ensemble average for large random graphs with specified degree distributions. This analysis is more generally applicable to the whole range of $\rho$. Numerical simulations reveal the fine structure of the distributions $\varphi$ and $\psi$. We are also able to compute a closed-form approximation for the large mean degree limit of Erd\H{o}s-R\'enyi random graphs.

\section{Message passing}
As detailed previously in \cite{Karrer2014}, analysis of the probability generating function of component sizes yields a set of self-consistency equations which can be solved efficiently by iteration. For a network with $M$ edges we define the $2M$-vector $\bm{H}$ to be the smallest solution in $[0,1]$ of the system 
\begin{equation}
H_{i\leftarrow j}=(1-\rho)+\rho\prod_{\ell\in\mathcal{N}_j\setminus i} H_{j\leftarrow \ell}\,.
\label{cavity}
\end{equation}
Here we write $\mathcal{N}_j$ for the neighbourhood of node $j$, and the entries of the vector $\bm{H}$ are indexed by ordered pairs of nodes attached by an edge. The expected size of the percolating cluster is then given by 
\begin{equation}
S=\frac{1}{N}\sum_{i=1}^N\left(1-\prod_{j\in\mathcal{N}_i} H_{i\leftarrow j}\right)\,.
\label{defS}
\end{equation}
Unpacking the sum above yields additional information about the likely outcomes of the percolation process for individual nodes, specifically
\begin{equation}
\langle \sigma_i\rangle=1-\prod_{j\in\mathcal{N}_i} H_{i\leftarrow j}\,.
\label{defsig}
\end{equation}
Similarly, for the expected sizes of small clusters one obtains 
\begin{equation}
\langle n_i\rangle=1+\sum_{j\in\mathcal{N}_i}\frac{H'_{i\leftarrow j}}{H_{i\leftarrow j}}\,,
\end{equation}
where $\bm{H'}$ solves
\begin{equation}
H_{i\leftarrow j}'=\rho\left(1+\sum_{\ell\in\mathcal{N}_j\setminus i}\frac{H'_{j\leftarrow \ell}}{H_{j\leftarrow \ell}}\right)\prod_{\ell\in\mathcal{N}_j\setminus i} H_{j\leftarrow \ell}\,.
\label{cavity_prime}
\end{equation}
For fixed finite graphs, both \eqref{cavity} and \eqref{cavity_prime} can be solved quickly by simple iteration, making analysis of percolation computationally tractable even in very large networks.

Notice that equation \eqref{cavity} admits a trivial fixed point in which $H_{i\leftarrow j}\equiv 1$, yielding $S=0$. This solution may or may not be stable, and the transition boundary exactly corresponds to the percolation threshold $\rho_c$. In the following section we examine in detail the behaviour of node-specific probability $\langle \sigma_i \rangle$ to belong to the giant cluster and the expected size $\langle n_i \rangle$ of finite clusters in the vicinity of the percolation phase transition. 
\section{Weakly non-linear analysis}
\label{weak}
\subsection{Percolation probability}
The value of the percolation threshold for a fixed finte network is revealed by linear stability analysis of the message passing equations. For a given ordered pair $i\leftarrow j$, the pair $k\leftarrow \ell$ will appear on the right hand side of \eqref{cavity} if and only $j=k$ and $\ell\neq i$. The matrix $B$ encoding this relationship is exactly the non-backtracking matrix mentioned in the introduction. The instability of the $H_{i\leftarrow j}\equiv 1$ solution is thus seen to occur at $\rho_c$ satisfying $\rho_c \lambda_{\rm max}(B)=1$.

Here we will consider $\rho=\rho_c+\varepsilon$ for small positive $\varepsilon$, and postulate a Taylor expansion for the $2M$-vector of messages
\begin{equation}
\bm{1}-\bm{H}=\varepsilon \bm{a}+\varepsilon^2\bm{b}+\varepsilon^3\bm{c}+\mathcal{O}(\varepsilon^4)\,,
\label{exp1-H}
\end{equation}
where $\bm{a}$, $\bm{b}$ and $\bm{c}$ are constant vectors, and $\bm{1}$ is the vector of ones, i.e. $\bm{1} =(1,1,\dots,1)^T$. Inserting into both sides of \eqref{cavity} and matching powers of $\varepsilon$ to third order we obtain the equations
\begin{equation}
\begin{split}
\bm{a}&=\rho_cB\bm{a}\,,\\
\bm{b}&=\rho_cB\bm{b}-\frac{\rho_c}{2}\Big((B\bm{a})^2-B\bm{a}^2\Big)+B\bm{a}\,,\\
\bm{c}&=\rho_cB\bm{c}-\rho_c\Big(B\bm{a}B\bm{b}-B(\bm{a}\bm{b})\Big)\\&\hspace{13mm}+\frac{\rho_c}{6}\Big((B\bm{a})^3-3B\bm{a}B\bm{a}^2+2B\bm{a}^3\Big)\\&\hspace{13mm}+B\bm{b}-\frac{1}{2}\Big((B\bm{a})^2-B\bm{a}^2\Big)\,.
\end{split}
\label{wnl}
\end{equation}
Here we use the notational shorthand of applying multiplication and exponentiation element-wise so that, for example, $\bm{a}\bm{b}$ denotes the vector with entries $a_{i\leftarrow j}b_{i\leftarrow j}$ and $\bm{a}^2$ that with entries $a_{i\leftarrow j}^2$. 

The first equation in \eqref{wnl} simply states that $\bm{a}$ is an eigenvector of $B$ associated to the Frobenius eigenvalue $\lambda_{\max}(B)=1/\rho_c$. To obtain the proper normalisation for $\bm{a}$ it is necessary to check the solvability of the second order equation. Let us write $\bm{a}=\alpha \bm{v}$, where $\bm{v}$ is the right Frobenius eigenvector of $B$ with positive entries, summing to one. We also introduce the corresponding left eigenvector $\bm{u}$, again with positive entries. That is,
\begin{equation}
\bm{v}/\rho_c=B\bm{v}\,,\quad\bm{u}^T/\rho_c=\bm{u}^TB\,,\quad \|\bm{v}\|_1=1\, .
\end{equation}
Following standard conventions by normalising left and right eigenvectors to form a bi-orthonormal system and, having fixed $\|\bm v\|=1$, we need to choose $\|\bm{u}\|_1$ such as to achieve $\bm{u}^T \bm{v}=1$. Then, replacing $\bm{a}$ by $\alpha\bm{v}$, the second order equation from (\ref{wnl}) reads as:
\begin{equation}
\bm{b}=\rho_cB\bm{b}-\alpha^2\frac{\bm{v}^2}{2\rho_c}+\alpha^2\frac{\rho_c}{2}B\bm{v}^2+\alpha\frac{\bm{v}}{\rho_c}\,.
\end{equation}
Multiplying through on the left by $\bm{u}^T2\rho_c/\alpha$ cancels some terms, yeilding $\alpha=2/\bm{u}^T\bm{v}^2(1-\rho_c)$. Returning to (\ref{defsig}), we find the first order approximation
\begin{equation}
\langle\sigma_i\rangle\approx\alpha(\rho-\rho_c)\sum_{j\in\mathcal{N}_i} v_{i\leftarrow j}\,.
\end{equation}
In fact the right hand side here is exactly the so-called ``non-backtracking centrality'' that was proposed in \cite{Martin2014}.
If the original system was simply connected to begin with, our choice of normalisation for $\bm v$ implies that $0 <v_{\rm min}\le v_{i\leftarrow j}\le v_{\rm max} <1$, entailing that close to the percolation threshold  we have lower and upper bounds for $\langle\sigma_i\rangle$ which are proportional to degree and of the form $\alpha(\rho-\rho_c)v_{\rm min} k_i \lesssim \langle\sigma_i\rangle \lesssim \alpha(\rho-\rho_c)v_{\rm max} k_i$. If the system was not simply connected to begin with, we have $v_{\rm min}=0$, and the lower bound becomes trivial.

By the same methodology we are also able to obtain the curvature of $\langle\sigma_i\rangle$ near criticality. The $\bm{b}$ equation from \eqref{wnl} can be rewritten as
\begin{equation}
(I-\rho_cB)\bm{b}=-\frac{\bm{a}^2}{2\rho_c}+\frac{\rho_cB\bm{a}^2}{2}+\frac{\bm{a}}{\rho_c}\,,
\end{equation}
and thus
\begin{equation}
\bm{b}=(I-\rho_cB)^+\left(\frac{\rho_cB\bm{a}^2}{2}-\frac{\bm{a}^2}{2\rho_c}+\frac{\bm{a}}{\rho_c}\right)+\beta {\bm v}\,,
\end{equation}
where $\,^+$ denotes the Moore-Penrose pseudoinverse and $\beta\in\mathbb{R}$ is a constant yet to be established. Introducing $\bm{w}=\bm{b}-\beta \bm{v}$, we multiply third order equation in \eqref{wnl} through on the left by $\bm{u}^T\rho_c$ to obtain
\begin{equation}
\begin{split}
0&=-\bm{u}^T\rho_c\Big(\bm{a}B\bm{w}-\bm{a}\bm{w}\Big)+\frac{1}{6\rho_c}\bm{u}^T\bm{a}^3-\frac{\rho_c}{2}\bm{u}^T(\bm{a}B\bm{a}^2)\\
&\quad+\frac{\rho_c}{3}\bm{u}^T\bm{a}^3+\bm{u}^T\bm{w}-\frac{1}{2\rho_c}\bm{u}^T\bm{a}^2+\frac{1}{2}\bm{u}^T\bm{a}^2\\
&\quad+\beta \left(\rho_c\bm{u}^T\bm{a}\bm{v}-\bm{u}^T \bm{a}\bm{v}+\bm{u}^T\bm{v}\right)\,.
\end{split}
\end{equation}
The constant $\beta$ is easily obtained by rearranging. 
\begin{figure}
\includegraphics[width=245pt, trim=10 10 10 10]{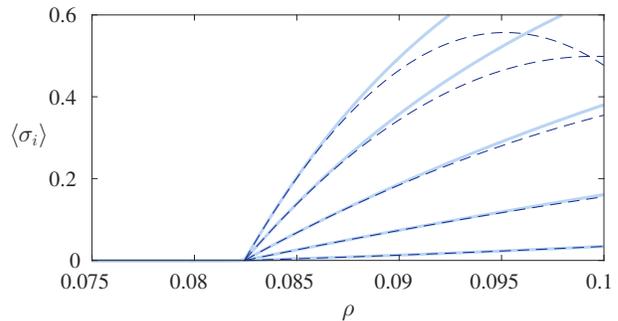}
\caption{Solid lines show the probability $\langle\sigma_i\rangle$ to appear in the percolating cluster, compared to the second order theory of Eq.~\eqref{sig2} in dashed lines. A selection of results are shown for vertices from a graph with power-law degree distribution, $p(k)\propto k^{-3}$ with $k_{\rm min}=2$.}
\label{WNLfig}
\end{figure}

Having solved for $\bm{a}$ and $\bm{b}$, a complete second order expansion for the probability of appearing in the percolating cluster is then given by
\begin{equation}
\begin{split}
\langle\sigma_i\rangle&\approx(\rho-\rho_c)\sum_{j\in\mathcal{N}_i} a_{i\leftarrow j}\\
&\quad+(\rho-\rho_c)^2\sum_{j\in\mathcal{N}_i}\left(b_{i\leftarrow j}-\frac{1}{2}a_{i\leftarrow j}\sum_{\ell\in\mathcal{N}_i\setminus j}a_{i\leftarrow \ell}\right)\\
\end{split}
\label{sig2}
\end{equation}
Figure~\ref{WNLfig} shows some numerical examples. 

\subsection{Small clusters}
Turning attention now to the expected size of finite clusters containing a particular node, we first note that below percolation we have $\bm{H}=\bm{1}$ and hence \eqref{cavity_prime} admits the exact solution 
\begin{equation}
\bm{H'}=\rho(I-\rho B)^{-1}\bm{1}\,.
\end{equation}
The matrix inverse here implies an order one pole at $\rho_c$, around which we develop a first order perturbation theory. If $\rho=\rho_c-\varepsilon$ and $\bm{H'}=\varepsilon^{-1}\bm{x}+\bm{y}+\mathcal{O}(\varepsilon)$ then
\begin{equation*}
\varepsilon^{-1}\bm{x}+\bm{y}+\mathcal{O}(\varepsilon)=(\rho_c-\varepsilon)(\bm{1}+\varepsilon^{-1}B\bm{x}+B\bm{y}+\mathcal{O}(\varepsilon))
\end{equation*}
Separating orders yields
\begin{equation}
\begin{split}
\bm{x}&=\rho_cB\bm{x}\\
\bm{y}&=\rho_cB\bm{y}+\rho_c\bm{1}-B\bm{x}\,.
\end{split}
\end{equation}
Evidently $\bm{x}=\xi^{(-)}\bm{v}$ for some constant $\xi^{(-)}$. To determine the constant $\xi^{(-)}$, we consult the second order equation, as usual multiplying on the left by $\bm{u}^T$, to find
\begin{equation}
\xi^{(-)}=\rho_c^2\|\bm u\|_1\,.
\end{equation}
To examine behaviour on the other side of the critical point, we set $\rho=\rho_c+\varepsilon$ and do the expansion again, this time using the result for 
$\alpha$ from the previous calculation.
\begin{equation}
\begin{split}
&\varepsilon^{-1}\bm{x}+\bm{y}+\mathcal{O}(\varepsilon)\\
&=(\rho_c+\varepsilon)\Big(\bm{1}+B(\varepsilon^{-1}\bm{x}+\bm{y}+\mathcal{O}(\varepsilon))(\bm{1}+\varepsilon\bm{a}+\mathcal{O}(\varepsilon^2))\Big)\\&\qquad\times\Big(\bm{1}-B(\varepsilon\bm{a}+\mathcal{O}(\varepsilon^2))\Big)\\
&=\varepsilon^{-1}\rho_cB\bm{x}+\rho_c\Big(\bm{1}+B(\bm{x}\bm{a})+B\bm{y}-B\bm{x}B\bm{a}\Big)\\&\qquad+B\bm{x}+\mathcal{O}(\varepsilon)\,.
\end{split}
\end{equation}
At leading order we again find a multiple of the Frobenius eigenvector, $\bm{x}=\xi^{(+)}\bm{v}$. The same trick of multiplying the second order term by the left eigenvector determines the constant:
\begin{equation}
\begin{split}
&0=\rho_c\|\bm u\|_1 +\alpha\xi^{(+)}\bm{u}^T\bm{v}^2-\frac{\alpha\xi^{(+)}}{\rho_c}\bm{u}^T\bm{v}^2+\frac{\xi^{(+)}}{\rho_c}\\
&\qquad\Rightarrow\quad \xi^{(+)}=\xi^{(-)}=\xi:=\rho_c^2 \|\bm u\|_1\,.
\end{split}
\end{equation}
We conclude that near percolation, typical size of finite clusters involving node $i$ is symmetric around $\rho_c$, with the asymptotic form 
\begin{equation}
\langle n_i\rangle = \frac{\rho_c^2\|\bm u\|_1}{|\rho-\rho_c|} \sum_{j\in\mathcal{N}_i}v_{i\leftarrow j}+\mathcal{O}(1)\,.
\label{n1}
\end{equation}
\begin{figure}
\includegraphics[width=245pt, trim=0 0 0 0]{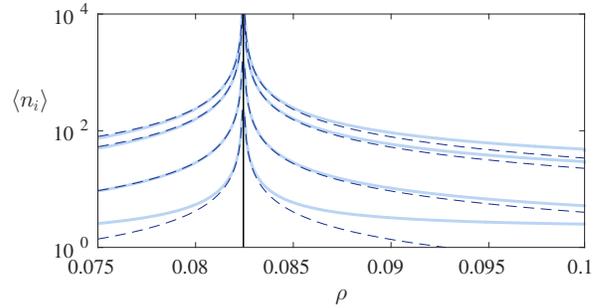}
\caption{Solid lines show expected size $\langle n_i\rangle$ of finite clusters containing given nodes, compared to the asymptotic theory of Eq.~\eqref{n1} in dashed lines. A selection of results are shown for vertices from a graph with power-law degree distribution, $p(k)\propto k^{-3}$ with $k_{\rm min}=2$.}
\label{WNLfig2}
\end{figure}
Note that this expression is again proportional to the non-backtracking centrality of node $i$. Figure~\ref{WNLfig2} shows some numerical examples. Note also that Eq. \eqref{n1} also allows to obtain an upper bound on the $\langle n_i\rangle$ in the vicinity of the percolation transition. As before, we can exploit the bounds $0 <v_{\rm min}\le  v_{i\leftarrow j}\le v_{\rm max} <1$ for systems which are origninally simply connected; they entail that in the vicinity of the percolation transition we have diverging lower and upper bounds for $\langle n_i\rangle$ of the form
\begin{equation}
\frac{\rho_c^2\|\bm u\|_1 v_{\rm min}}{|\rho-\rho_c|}\, k_i \lesssim \langle n_i\rangle \lesssim \frac{\rho_c^2\|\bm u\|_1 v_{\rm max}}{|\rho-\rho_c|}\, k_i \, .
\label{ni}
\end{equation}
Upper and lower bounds for the $\langle n_i\rangle$ therefore proportional to the degree $k_i$ of the site in question, with the constant of proportionality diverging as the transition is approached. As for the $\langle \sigma_i\rangle$ the lower bound becomes trivial for systems which are not simply connected to begin with.

\section{Population dynamics for $N\to \infty$}
\label{popdyn}
\subsection{Percolation probability}
We now proceed to analyse the bond percolation problem in the thermodynamic limit $N\to \infty$ for random networks in the configuration model class. In this limit Eqs.~\eqref{cavity} can be interpreted as a stochastic recursion for the values of $H_{i\leftarrow j}$ of randomly chosen neighbouring nodes $i$ and $j$. A probability density function $\widehat\varphi$ for the values of $H_{i\leftarrow j}$ will then have to satisfy a self-consistency equation of the form
\begin{equation}
\widehat\varphi(h)=\sum_{k\geq 1}\frac{kp(k)}{c}\int \prod_{\ell=1}^{k-1} \text{d} \widehat\varphi(h_\ell)\, \delta\left(h-1+\rho-\rho\prod_{\ell=1}^{k-1} h_\ell\right)
\label{RDE}
\end{equation}
in which $p(k)$ is the degree distribution, and $\text{d} \widehat\varphi(h_\ell)$ is shorthand for $ \text{d}h_\ell \widehat\varphi(h_\ell)$. It is obtained by averaging the r.h.s of \eqref{cavity} over all realisations for which $H_{i\leftarrow j}\in [h,h+\text{d}h]$.
The factor of $kp(k)/c$ above is the branching distribution of the configuration model, giving the distribution of the number of additional edges incident upon node $j$ from the randomly chosen pair $i\leftarrow j$. Equation~\eqref{RDE} simply states that if the $H_{j\leftarrow\ell}$ appearing on the right hand side of \eqref{cavity} are random variables with distribution $\widehat\varphi$, then across neighbourhoods $\mathcal{N}_j$ the resulting $H_{i\leftarrow j}$ should have the same distribution.  

The distribution $\varphi$ of the probabilities to remain in the percolating cluster is expressed in terms of the solution $\widehat\varphi$ of Eq. \eqref{RDE} via
\begin{equation}
\varphi(s)=\sum_{k\geq 0}p(k)\int \prod_{\ell=1}^{k}\text{d}\widehat\varphi(h_\ell)\,\delta\left(s-1+\prod_{\ell=1}^{k} h_\ell\right).
\label{PD_phi}
\end{equation}

\begin{figure}
\includegraphics[width=245pt, trim=10 10 10 10]{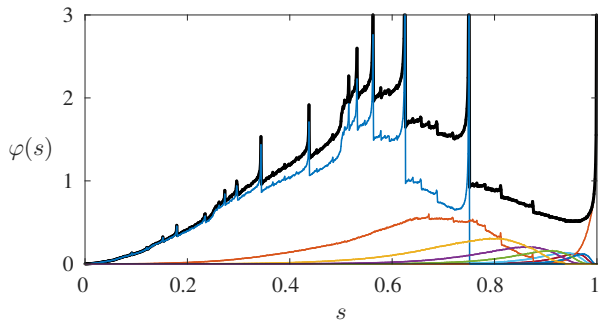}
\caption{Distribution $\varphi(s)$ of probabilities to be part of the giant cluster for the percolation problem on a graph with power-law degree distribution, $p(k)\propto k^{-3}$ with $k_{\rm min}=2$, at $\rho=0.5$ (black). Results of population dynamics shown together with its unfolding according to degree for $k=2,3,\dots,9$, and $\{k\ge 10\}$ (blue, red, green,\dots from left to right).}
\label{PD}
\end{figure}

Equation \eqref{RDE} does not offer much hope for exact solution, however, highly accurate numerical solutions can be obtained by iterating a large sample population $\{h_\omega\}_{\omega=1}^\Omega$, a technique known in this context as population dynamics \cite{Mezard2001}. In Figure~\ref{PD} we show the results of the population dynamics algorithm for random graphs with a power-law degree distributions of the form $p(k) \propto k^{-3}$, for $k\ge k_{\rm min}=2$, and a edge occupation probability of $\rho=1/2$. Also shown is a partial unfolding of $\varphi(s)$ according to degree $k$. The distributions show a significant amount of structure, including some sharply peaked `band-edges' of a type that also appear in the Gnutella data shown in Fig. \ref{gnutella} for sufficiently large $\rho$. 

More insight into the structure of the distribution $\varphi$ can be gained by disentangling the contributions coming from nodes of different degrees. We decompose \eqref{PD_phi} into $\varphi(s)=\sum_{k\geq 0}p(k)\varphi_k(s)$, where 
\begin{equation}
\begin{split}
\varphi_k(s)&=\int \prod_{\ell=1}^{k}\text{d}\widehat\varphi(h_\ell)\, \delta\left(s-1+\prod_{\ell=1}^{k} h_\ell\right)\,.
\end{split}
\label{phi-dec}
\end{equation}
This unfolding according to degree is also shown in Figure~\ref{PD}. It reveals that -- to the resolution shown -- the peaked band-edges mentioned above are associated with degree 2 and degree 3 vertices, with the exception of the peak at $s=1$ which originates from $\varphi_{k\ge 10}(s)$, i.e. from large-degree vertices. Moreover, for each degree $k$, there is an upper cutoff $s_k <1$ of the probability, beyond which $\varphi_k(s)=0$. A bound for this cutoff is easily read off from the single instance equations. Indeed Eq. \eqref{cavity}  implies that $H_{i\leftarrow j} \ge 1-\rho$, which via Eq. \ref{defsig} in turn entails that $\langle \sigma_i\rangle \le 1-(1-\rho)^{k_i}$. Hence we have the bounds $s_k \le 1-(1-\rho)^k$ for the $k$-dependent cutoffs. Numerical evidence suggests that these bounds are rather sharp. Their exact form is very likely amenable to a more detailed probabilistic analysis. 

\subsection{Small clusters}
The population dynamics approach is also available for the distribution of expected finite cluster sizes in the thermodynamic limit. From Eqs. \eqref{cavity} and \eqref{cavity_prime} we derive a recursion equation for the joint distribution $\widehat\psi(h,h')$ of messages $H$ and $H'$,
\begin{equation*}
\begin{split}
&\widehat\psi(h,h')=\\&\quad\sum_{k\geq 1}\frac{kp(k)}{c}\int \prod_{\ell=1}^{k-1}\text{d}\widehat\psi(h_\ell,h'_\ell)\,\delta\left(h-1+\rho-\rho\prod_{\ell=1}^{k-1} h_\ell\right)\\&\hspace{35mm}\delta\left(h'-\rho\left(1+\sum_{\ell=1}^{k-1}\frac{h'_\ell}{h_\ell}\right)\prod_{\ell=1}^{k-1} h_\ell\right)\, ,
\label{RDE2}
\end{split}
\end{equation*}
in which we use an analogous shorthand for integration measures, $\text{d}\widehat\psi(h_\ell,h'_\ell)\equiv \text{d}h_\ell \text{d}h'_\ell\, \widehat\psi(h_\ell,h'_\ell)$. 
This equation generalises \eqref{RDE}, as $\widehat\varphi$ is recovered as the marginal $\widehat\varphi(h)=\int  \text{d}h'\,\widehat\psi(h,h')$. The distribution of finite cluster sizes is then given by $\psi(n)=\sum_{k\geq 0}p(k)\psi_k(n)$, where 
\begin{equation}
\begin{split}
\psi_k(n)&=\int \prod_{\ell=1}^{k}\text{d}\widehat\psi(h_\ell,h'_\ell)\,\delta\left(n-\left(1+\sum_{\ell=1}^{k}\frac{h'_\ell}{h_\ell}\right)\right)
\end{split}
\end{equation}
\begin{figure}
\includegraphics[width=245pt, trim=10 10 10 10]{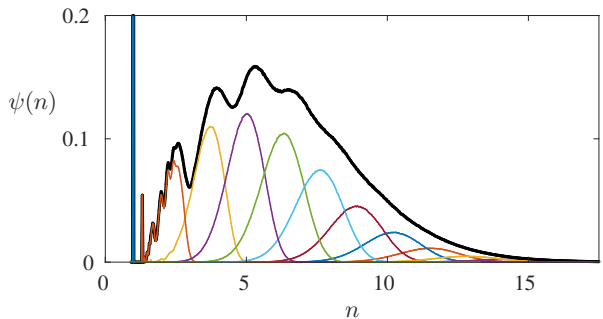}
\caption{Distribution $\psi(n)$ of average cluster sizes for percolation on an Erd\H{o}s-R\'enyi network of mean degree $c=4$ at $\rho=0.3$. Results of population dynamics shown together with its unfolding according to degree for $k=0,1, 2,\dots,9$, and $\{k\ge 10\}$ (blue, red, green,\dots from left to right).}
\label{PD2}
\end{figure}

Figure.~\ref{PD2} shows the results of numerical solution of the population dynamics equations for $\psi(n)$ in the case of a Poisson random graph of mean degree $c=4$ at bond occupation probability $\rho=0.3$, together with a deconvolution according to degree $k$. 

The figure exhibits two $\delta$-peaks at $n=1$ and $n=1+\rho=1.3$, the first corresponding to isolated sites with $k_i=0$, which remain isolated after deleting bonds, the second giving the average cluster size of pairs of vertices, which formed isolated dimers in the original graph (before deleting bonds), and consequently appears in $\psi_1(n)$. There are in principle many more such $\delta$-peaks related to the contribution of originally isolated clusters to the $\langle n_i\rangle$-statistics. For the present system their weight is, however, too small to be detected in the continuum at the resolution chosen. The continuous part of $\psi(n)$ reveals several discernible peaks, which according to the deconvolution can be associated with vertices of degrees $k=1, 2, 3$, and 4, whereas the peaks of $\psi_k(n)$ for larger $k$ are too narrowly separated compared to their widths to give rise to discernible features in the overall distribution $\psi(n)$ at larger $n$. Each of the degree dependent $\psi_k(n)$ also exhibits an upper cutoff $n_k$ beyond which $\psi_k(n) = 0$. 

\subsection{Large mean degree}
Further analytical progress can be made in the limit of large mean degree and small edge occupancy; here we develop a single defect approximation \cite{Biroli1999} for Poisson random graphs. In this approximation scheme we assume that $\widehat\varphi(h)$ is for large mean degree $c$ well approximated by a a Dirac delta distribution, $\widehat\varphi(h)=\delta(h-h_\star)$. Inserting this ansatz into \eqref{RDE} and integrating over $h$ we obtain $h_\star=1-\rho+\rho e^{-c(1-h_\star)}$, from which it follows that 
\begin{equation}
\quad h_\star=\begin{cases}1-\rho-\frac{1}{c}W\big(-c\rho e^{-c\rho}\big)\quad&\rho\geq\rho_c\\1&\rho<\rho_c\,,\end{cases}
\end{equation}
where $W$ is the Lambert W function, and the percolation transition occurs at $\rho_c=1/c$. To obtain non-trivial behaviour in the limit $c\to\infty$, it is therefore necessary to introduce the scaling $\rho=\varrho/c$ where $\varrho>1$. Then $h_\star=1-\eta/c$, where $\eta=\varrho+W(-\varrho e^{-\varrho})$.

\begin{figure}
\includegraphics[width=245pt, trim=10 10 10 10]{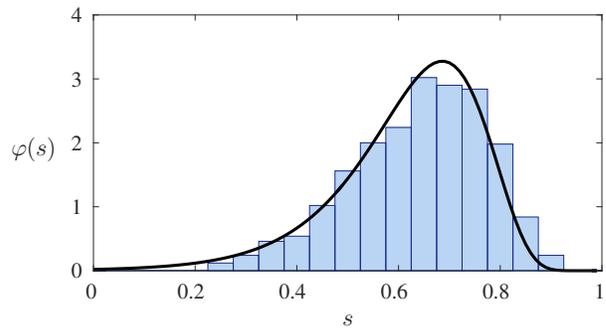}
\caption{Comparison of the large $c$ asymptotic given in Eq.~\eqref{largec}, with a histogram taken from a finite single instance of a Poisson graph with $N=1000$ nodes and mean degree $c=8$. The edge occupancy is $\rho=0.2$.}
\label{c8}
\end{figure}

Imagining a single ``defect'' node $i$ with degree $k_i$ attached to the otherwise homogeneous network, we find the probability of this node to appear in the percolating cluster to be 
\begin{equation}
\langle \sigma_i\rangle=1-h_\star^{k_i}\approx1-e^{-\eta k_i/c}\,.
\end{equation}
Taking the Gaussian limit of the Poisson distribution at large mean degree $c$, we note that the distribution of the ratio $k_i/c$ of a randomly selected node $i$ is well-approximated by a normal random variable $x$ with mean one and variance $1/c$. Under these assumptions we change probability variables from $x$ to $s$, computing
\begin{equation}
\varphi(s)\approx\left|\frac{d}{ds}f^{-1}(s)\right|\sqrt{\frac{c}{2\pi}}\exp\left\{-\frac{c}{2}\big(1-f^{-1}(s)\big)^2\right\}\,,
\end{equation}
where $f(x)=1-e^{-\eta x}.$ The inverse and its derivative are
\begin{equation}
f^{-1}(s)=-\frac{\log(1-s)}{\eta}\,,\quad\frac{d}{ds}f^{-1}(s)=\frac{1}{\eta (1-s)}.
\end{equation}
Putting all this together we arrive at 
\begin{equation}
\varphi(s)\approx\frac{\exp\left\{-\frac{c}{2}\left(1+\frac{1}{\eta}\log(1-s)\right)^2-\log(1-s)\right\}}{\sqrt{2\pi/c}\,\eta}\,.
\label{largec}
\end{equation}
Although it is exact only in the limit of large $c$, this approximation holds remarkably well for smaller values; see Fig.~\ref{c8} for an example with $c=8$. As $c\to\infty$, the distribution $\varphi$ approaches a Gaussian with mean $1-e^{-\eta}$ and variance $\eta^2e^{-2\eta}/c$.

Applying the same approach to $\psi$, the distribution of finite clusters, we obtain a simple Gaussian law 
\begin{equation}
\psi(n)\approx \frac{\sqrt{c}}{\sqrt{2\pi}\gamma}\exp\left\{-\frac{c}{2\gamma^2}\left(n-1-\gamma\right)^2\right\}\,,
\end{equation}
where $\gamma=\varrho/(e^{\varrho+W(-\varrho e^{-\varrho})}-\varrho)$.

\section{Summary and Discussion}
To summarise, we have taken a new look at the message passing approach \cite{Karrer2010, Karrer2014} to bond percolation on complex networks, revealing a considerable degree of heterogeneity. We have seen that this approach allows one to determine the distribution of probabilities of individual nodes to belong to the percolating cluster, as well as the distribution of the average sizes of non-percolating clusters to which individual nodes may belong. We found both distributions to be typically broad, so that the \textit{average} percolation probabilities and \textit{average} sizes of finite clusters that are typically reported in analyses of percolation on random networks must be regarded as poor representations of the actual heterogeneity that is present in this problem. We have also obtained deconvolution of both distributions according to degree, in analogy to the  deconvolution of sparse random matrix spectra advocated  in \cite{Kuehn2008}, and found that the component distributions $\varphi_k$ and $\psi_k$ are themselves non-degenerate (except at $k=0$) and indeed typically broad as well. A fairly detailed analysis of node-specific
percolation probabilities $\langle \sigma_i\rangle $ and average cluster-sizes $\langle n_i\rangle$ was provided in the vicinity of the percolation transition, and formulated in terms of spectral properties of the Hashimoto matrix $B$, notably the right and left eigenvectors corresponding to its largest eigenvalue. 

In the present paper we have used the message passing approach to analyse a specific finite large real world instance, percolation on the network of the Gnutella file sharing platform \cite{Snap2014}, and presented methods to analyse the problem for random graphs in the configuration model class in the thermodynamic limit of infinite system size, $N\to \infty$. We have presented examples for the distribution of percolation probabilities on a scale free graph with power-law degree distribution, and for the distribution of finite cluster sizes in the case of an Erd\H{o}s-R\'enyi network. These two examples can only scratch the surface of the variability of phenomena that might be observed. A few general trends may be noted though. If the original random network ensemble contains finite clusters to begin with, as is indeed the case for  Erd\H{o}s-R\'enyi networks of finite mean degree, then the resulting distribution
$\psi(n)$ of the average finite cluster sizes will contain a family of $\delta$-peaks originating from the distribution of the resulting
finite cluster sizes generated by bond removal, whereas the broad continuum is generated by clusters obtained as a result of disconnecting a finite connected set of vertices from the original percolating cluster. Only if the non-percolating fraction in the original graph is sufficiently large will the delta-peaks carry sufficient weight to be detectable in the population dynamics. This is in particular the case for low mean degree Erd\H{o}s-R\'enyi graphs. Cluster size-distributions typically get broader as the percolation transition is approached from above or below; the same is true after deconvolution meaning that peaks contributed by individual $\psi_k$ will cease to be discernible in the sum. The same broad trends are observed in $\varphi(s)$ and it's deconvolutions.

We have been able to obtain closed form analytic approximations for the distribution of percolation probabilities and mean cluster sizes in the large mean connectivity limit of Erd\H{o}-Renyi graphs, which produces excellent results already for fairly moderate values of $c$. The same methodology could be adapted to other degree distributions, and should be efficient whenever these distributions become narrow in the large mean degree limit. 

Returning to the heterogeneity of percolation probabilities across a network, and the practical relevance of this phenomenon in the context of cascading failures, epidemic spreading, or probabilistic information spreading, we note that the considerable detail which our methods allow to unearth might be useful for instance in the design of optimal vaccination strategies that exploit information beyond degree. Indeed the fact that percolation probabilities conditioned on degrees are themselves broadly distributed is a clear indicator of the fact that the degree based approximation on which the majority of attempts to design optimal vaccination strategies has been based misses important information which could be exploited to improve upon such strategies. We believe that this point is worth investigating.

\textit{Acknowledgements}\\
TR is supported by the Royal Society.

\bibliography{HeteroPerc}

\end{document}